\documentstyle[12pt]{article}
\input{psfig.sty}

\begin{document}

\begin{titlepage}

\baselineskip 24pt

\begin{center}

{\Large {\bf A Dynamical Mechanism for Quark Mixing and Neutrino
   Oscillations}}\\
\vspace{1cm}

\baselineskip 16pt

{\large Jos\'e BORDES}\\
jose.m.bordes\,@\,uv.es\\
{\it Dept. Fisica Teorica, Univ. de Valencia,\\
  c. Dr. Moliner 50, E-46100 Burjassot (Valencia), Spain}\\
\vspace{.2cm}
{\large CHAN Hong-Mo}\\
chanhm\,@\,v2.rl.ac.uk \\
{\it Rutherford Appleton Laboratory,\\
  Chilton, Didcot, Oxon OX11 0QX, United Kingdom}\\
\vspace{.2cm}
{\large TSOU Sheung Tsun}\\
tsou\,@\,maths.ox.ac.uk\\
{\it Mathematical Institute, University of Oxford,\\
  24-29 St. Giles', Oxford OX1 3LB, United Kingdom}

\end{center}

\baselineskip 14pt

\begin{abstract}
We show that assuming fermion generations to be given by a gauge
symmetry plus a certain Higgs mechanism for its breaking, the known
empirical features of quark and lepton mixing can be largely explained,
including in particular the fact that the mixing (CKM) matrix element 
$U_{\mu3}$ responsible for the muon anomaly in atmospheric neutrinos is 
near maximal and much larger than their quark counterparts $V_{cb}$ and
$ V_{ts}$, while the corner elements for both quarks ($V_{ub}, V_{td}$)
and leptons ($U_{e3}$) are all very small.  The mechanism also gives 
automatically a hierarchical fermion mass spectrum which is intimately 
related to the mixing pattern.

\end{abstract}

\end{titlepage}

\clearpage

The quark mixing pattern as measured by the Cabibbo-Kobayashi-Moskawa
(CKM) matrix is now quite well-known.  The latest databook \cite{databook}
gives the absolute values of the matrix elements as:
\begin{eqnarray}
\lefteqn{
\left( \begin{array}{ccc} |V_{ud}| & |V_{us}| & |V_{ub}| \\
                          |V_{cd}| & |V_{cs}| & |V_{cb}| \\
                          |V_{td}| & |V_{ts}| & |V_{tb}|  
   \end{array} \right)
   =} \nonumber\\
&& \left( \begin{array}{lll} 
   0.9745-0.9760 & 0.217-0.224 & 0.0018-0.0045 \\
   0.217-0.224 & 0.9737-0.9753 & 0.036-0.042 \\
   0.004-0.013 & 0.035-0.042 & 0.9991-0.9994  \end{array} \right).
\label{exckmq}
\end{eqnarray}
Information on the corresponding matrix for leptons is beginning also to 
emerge from recent experiments on neutrino oscillations.  In particular, the 
result from atmospheric neutrinos \cite{kamioka,superk,soudan,newSK1} 
shows that the mixing angle between $\nu_\mu$ and the heaviest $\nu_3$ 
state is near maximal, while the absence of oscillation effects in some 
reactor experiments, in particular CHOOZ \cite{chooz}, implies that the 
mixing of $\nu_e$ to the same heaviest state $\nu_3$ is rather small.  From 
solar neutrino data, the picture is not yet entirely clear.  Of the 3 
traditional solutions, namely (i) the small angle MSW, (ii) the large angle 
MSW, and (iii) the long wave-length (or vacuum, or just-so) oscillation 
(LWO), both (i) and (ii) are under pressure from the latest Superkamiokande 
data on day-night variation and flux \cite{newSK2}, which seem to have 
a slight preference for (iii), but the situation is still far from settled.
On can conclude at present only that the angle between $\nu_e$ and the
second heaviest state $\nu_2$ is either quite small (i) or again near 
maximal (ii)--(iii).  As a result, a CKM matrix is suggested roughly of 
the form:
\begin{equation}
\left( \begin{array}{ccc} |U_{e1}| & |U_{e2}| & |U_{e3}| \\
                          |U_{\mu1}| & |U_{\mu2}| & |U_{\mu3}| \\
                          |U_{\tau1}| & |U_{\tau2}| & |U_{\tau3}|
   \end{array} \right)
   = \left( \begin{array}{ccc} \star & 0.4-0.7 & 0.0-0.15 \\
                               \star & \star & 0.56-0.83 \\
                               \star & \star & \star \end{array} \right),
\label{exckml}
\end{equation}
where, for reasons which will be apparent later, we have inserted for 
$U_{e2}$ the value suggested by the LWO solution (iii).  If CP-violations 
are ignored, the elements denoted by $\star$ are obtainable by unitarity 
from the others.

In these mixing matrices, one notices some very outstanding features:
\begin{description}
\item{(a)} The off-diagonal elements in the quark CKM matrix are all
small or very small;
\item{(b)} The corner elements in both the quark and lepton matrices
are all very much smaller than the others;
\item{(c)} The $U_{\mu3}$ element in the lepton matrix is much (about
a factor 20) larger than its quark counterparts, namely $V_{cb}$ and
$V_{ts}$.
\end{description}
These features, together with the actual values that the elements take,
cry out urgently for a theoretical explanation.

What we wish to show in this paper is that all the above features
together with the hierarchical fermion mass spectrum can very simply 
be explained and even semi-quantitatively calculated in terms of a few
parameters if one assumes generation to be an $SU(3)$ gauge symmetry
spontaneously broken in a particular manner.  This observation is
abstracted from a recently proposed scheme we called the Dualized
Standard Model (DSM) \cite{Vancouver,dualcons} based on a nonabelian 
generalization of electric-magnetic duality \cite{ymduality}.  Here we 
shall do the following.  First, we shall distill and simplify the 
arguments to such an extent as to make the mechanism, we hope, completely
transparent.  Secondly, we shall make clear that the main mechanism
is independent of the concept of duality, thus freeing it from our
own theoretical bias, so that if one prefers (which we ourselves do not 
for reasons to be given later) one can obtain similar results by grafting 
the proposed mechanism on to some different, not necessarily dual, scheme. 
Thirdly, we shall present a new, more systematic, fit together with a 
more detailed comparison with experiment using the latest data while 
making some points of detail not noted before.  

The idea that generation originates from a (spontaneously broken) 
`horizontal' gauge symmetry is not new.  The empirical fact that fermions 
seem to occur in 3 and only 3 generations suggests $SU(3)$.  In analogy
to the electroweak theory, we then propose to assign left-handed fermions
to the fundamental triplet representation and right-handed fermions to
singlets.  For breaking the symmetry, a possibility is to introduce 
3 $SU(3)$ triplets of Higgs fields, say $\phi^{(a)}, a = 1, 2, 3$, with 
linearly independent, say mutually orthogonal, vacuum values, namely
that $\bar{\phi}^{(a)} \phi^{(b)} = 0, a \neq b$ at vacuum.  Furthermore,
we stipulate that the 3 Higgs triplets be `indistinguishable' so that 
the action has to be symmetric under their permutations, although the 
vacuum need not be thus symmetric.\footnote{In the DSM, these proposals 
are given some {\it raison d'\^etre} since there the $\phi$'s are related 
to frame vectors in $U(3)$, but one need take no account of that if one 
so prefers.}  

A possible potential for these Higgs fields is then:
\begin{equation}
V[\phi] = -\mu \sum_{(a)} |\phi^{(a)}|^2 + \lambda \left\{ \sum_{(a)}
   |\phi^{(a)}|^2 \right\}^2 + \kappa \sum_{(a) \neq (b)} |\bar{\phi}^{(a)}
   .\phi^{(b)}|^2,
\label{Vofphi}
\end{equation}
for which a general vacuum can be expressed as:
\begin{equation}
\phi^{(1)} = \zeta \left( \begin{array}{c} x \\ 0 \\ 0 \end{array} \right);
\phi^{(2)} = \zeta \left( \begin{array}{c} 0 \\ y \\ 0 \end{array} \right);
\phi^{(3)} = \zeta \left( \begin{array}{c} 0 \\ 0 \\ z \end{array} \right),
\label{phivac}
\end{equation}
with 
\begin{equation}
\zeta = \sqrt{\mu/2\lambda},
\label{zeta}
\end{equation}
and $x, y, z$ all real and positive, satisfying:
\begin{equation}
x^2 + y^2 + z^2 = 1.
\label{xyznorm}
\end{equation}
Such a vacuum breaks the permutation symmetry of the $\phi$'s, and also
the $SU(3)$ gauge symmetry completely.  As a result, all the vector gauge 
bosons in the theory acquire a mass, eating up all but 9 of the Higgs
modes\footnote{The $\phi$'s in fact break a larger $U(3)$ symmetry,
giving thus 9 massive vector bosons.}.

Next, given the above assignments of $SU(3)$ representations to the left- 
and right-handed fermions, the Yukawa couplings take the form:
\begin{equation}
\sum_{(a)[b]} Y_{[b]} \bar{\psi}_L^a  \phi^{(a)}_a \psi_R^{[b]},
\label{Yukawa}
\end{equation}
which is symmetric under permutations of $\phi^{(a)}$ as required.  As a 
result, the tree-level mass matrix for each of the 4 fermion-types $T$ 
(i.e. whether $U$- or $D$-type quarks, or charged leptons (L) or neutrinos 
(N)) is of the following factorized form:
\begin{equation}
m \propto \left( \begin{array}{c} x \\ y \\ z \end{array} \right)
   (a, b, c),
\label{mtree}
\end{equation}
with $a, b, c$ being the Yukawa couplings $Y_{[b]}$.  Of more relevance
to the mass spectrum is the matrix $m m^{\dagger}$ which takes the form:
\begin{equation}
\sqrt{m m^\dagger} = m_T \left( \begin{array}{c} x \\ y \\ z 
   \end{array} \right) (x, y, z).
\label{mtreew}
\end{equation}
This is of rank 1, having only one nonzero eigenvalue with eigenvector 
$(x, y, z)$ the components of which, being Higgs vev's, are independent of 
the fermion-type $T$.  Hence we have already at tree-level (i) that the 
fermion mass spectrum is `hierarchical' with one generation much heavier 
than the other two, (ii) that the CKM matrix giving the relative orientation 
between the eigenvectors of the up- and down-type fermions is the identity 
matrix.  Both of these conditions give sensible zero-order approximations,
at least for quarks, to the experimental data.

Consider next 1-loop corrections.  It is not hard to see that the corrected
fermion mass matrix $m'$ will remain in a factorized form.  The reason is 
that only those loops involving the generation-changing gauge and Higgs
bosons can affect the factorization, and of these the gauge bosons couple 
only to the left-handed fermions while the Higgs bosons have couplings
which are themselves factorizable.  Indeed, it appears that the factorized
mass matrix will survive to all orders in perturbation.  As a result, we
have:
\begin{equation}
\sqrt{m' m'^\dagger} = m'_T \left( \begin{array}{c} x' \\ y' \\ z' 
   \end{array} \right) (x', y', z'),
\label{mloopw}
\end{equation}
where the corrected vector $(x', y', z')$ depends both on the fermion-type 
and on the energy scale.  At the 1-loop level, the vector $(x', y', z')$
remains real so that there is no $CP$-violation at this level.

The scale-dependence of $m'$ above is a special case of a mass matrix
which rotates with the energy scale.  In itself, this is not unusual since
already in the standard formulation of the Standard Model, such a rotation
of the fermion mass matrix will result in the renormalization group 
equation from a nondiagonal CKM matrix \cite{examp}, although the
effect there is small and therefore usually neglected.  When the effect
of the rotation is appreciable, as it can be in our present case, then 
care has to be exercised in its physical interpretation.  When the mass 
matrix does not rotate with scale, as in QCD where the scale-dependence
induced by gluonic corrections appears as an overall flavour-independent
factor, there is of course no difficulty in identifying the masses and 
state vectors of the physical states.  The matrix can be diagonalized 
at any scale giving a set of eigenvectors independent of the scale 
although the eigenvalues themselves will in general be scale-dependent.  
These eigenvectors can then be taken unambiguously as the state vectors 
of the physical states while the mass of each physical state can be 
defined as the running eigenvalue $m_i(\mu)$ corresponding to the state 
$i$ when taken at the scale equal to its value, namely as the solution 
to the equations $m_i(\mu) = \mu$.  However, if the mass matrix rotates 
with scale, then its eigenvectors are also scale-dependent and it becomes 
unclear how the physical state vectors are to be defined.  One may be 
tempted to define the eigenvector for the value $m_i$ at the scale $\mu$ 
satisfying the equation $m_i(\mu) = \mu$ as the state vector for the
physical state $i$, but the state vectors defined in this way will not 
be mutually orthogonal, thus contradicting the ansatz that they represent
physically independent quantum states.  

The solution we propose to adopt in this paper, which is in fact the only 
one we can think of, is as follows.  We run the mass matrix $m$ down in scale 
until we have for its highest eigenvalue $m_3$ a solution to the equation 
$m_3(\mu) = \mu$.  This value at this scale we define as the mass $m_3$, 
and the corresponding eigenvector the state vector ${\bf v_3}$ of the 
heaviest generation.  Below that energy, the state $3$ no longer exists 
as a physical state, and only the two lighter generations survive, the 
state vectors of which have to be orthogonal to ${\bf v_3}$.  We define 
therefore the mass matrix at energies below $m_3$ as the $2 \times 2$ 
submatrix $\hat{m}$ of $m$ in the subspace orthogonal to ${\bf v_3}$.  
To find now the mass and state vector for generation $2$ we follow 
with $\hat{m}$ the same procedure as for $3$ with $m$ and run $\hat{m}$ 
down in scale until we find a solution to the equation $\hat{m}_2(\mu) 
= \mu$, which value we call the mass $m_2$ and the corresponding eigenvector 
at that scale the state vector ${\bf v_2}$ of the generation $2$.  The 
state vector of the lightest generation $1$ is now also defined, as the 
vector orthogonal to both ${\bf v_3}$ and ${\bf v_2}$, while the mass of 
$1$ will obtain by repeating the above procedure, namely by running down
in scale the expectation value $\langle {\bf v_1}|m|{\bf v_1} \rangle$
until its value equals the scale.  In this way, each mass is evaluated at 
its own appropriate scale while the physical state vectors of the 3 
generations are all mutually orthogonal, as they should be.

\begin{figure}[htb]
\centerline{\psfig{figure=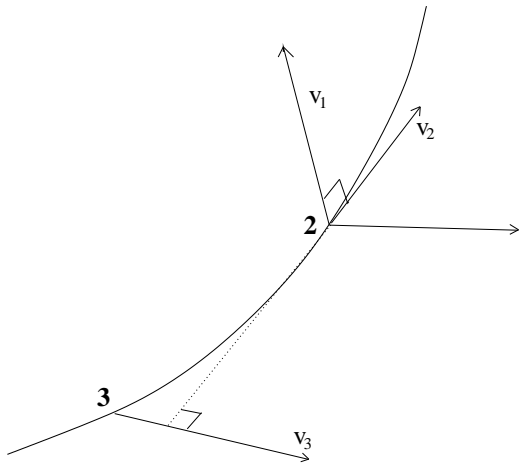,width=0.5\textwidth}}
\caption{The triad of state vectors for the 3 generations of fermions.}
\label{triad}
\end{figure}

Applying the above procedure to the factorized mass matrix $m'$ in 
(\ref{mloopw}), one sees that for the heaviest generation fermion of 
type $T$, the mass $m_3$ is $m'_T$ and the state vector ${\bf v_3}$  is
$(x', y', z')$, both taken at the scale $\mu$ satisfying the condition
$m'_T(\mu) = \mu$.  At that scale, the subspace orthogonal to ${\bf v_3}$ 
has zero mass eigenvalues, and it is as yet unclear which vector in 
it should correspond to the second and which the lightest generation.  
However, as the scale lowers further, the vector $(x', y', z')$ rotates 
to a different direction giving nonzero components in the orthogonal 
subspace and hence a nonzero eigenvalue to $\hat{m}'$.  One can then 
define this nonzero value as $\hat{m}'_2(\mu)$ and procede as above to 
determine the (nonzero) mass $m_2$ and state vector ${\bf v_2}$ of the
second generation.  At the same time one determines the state vector
${\bf v_1}$ of the lightest generation.  The triad of state vectors so
determined for the 3 generations are as shown in Figure \ref{triad}.
The mass of the lightest generation can also be found by running the 
scale down further.  As a result, all 3 generations will acquire finite 
masses by this `leakage' mechanism, but the mass spectrum will be 
hierarchical, meaning that $m_3 \gg m_2 \gg m_1$, qualitatively as 
experimentally observed.  Further, since a triad of state vectors for 
the 3 generations have now been defined for each fermion-type, CKM 
matrix elements can be evaluated as the direction cosines between the 
state vectors of the various up- and down-type fermions.  And since the 
loop corrections are in general different for up- and down-types, the 
resulting matrix will be nondiagonal giving nonzero mixing.

One sees therefore that in the present framework with a factorized mass 
matrix, nearly all the information on fermion mixing and much of that on 
the fermion mass spectrum are encoded in a single 3-vector $(x', y', z')$ 
in generation space, one for each fermion-type.  This vector rotates
with the energy scale and as the scale changes, it traces out a trajectory
on the unit sphere.  By studying the shape of these trajectories and
the speed at which $(x', y', z')$ moves along them, one will be able to 
deduce properties of the CKM matrix and the fermion mass spectrum.

\begin{figure}[htb]
\centerline{\psfig{figure=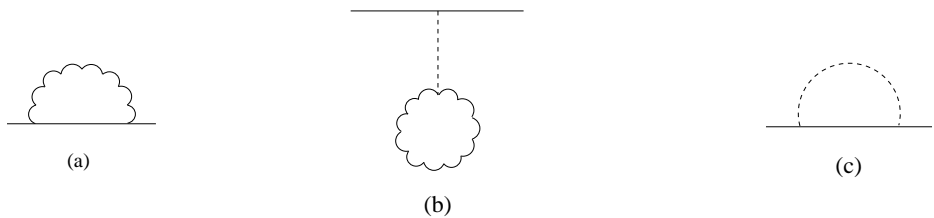,width=0.9\textwidth}}
\caption{One loop diagrams rotating the fermion mass matrix.}
\label{1loopdiag}
\end{figure}

Let us then examine in more detail how loop corrections affect the vector
$(x', y', z')$.  As already noted, only those loop diagrams involving
generation-changing bosons can rotate the vector $(x', y', z')$.  A closer
examination then reveals \cite{ourCKM} that of the various 1-loop diagrams, 
only 3 give rotations, namely those in Figure \ref{1loopdiag}, where a
full line denotes a fermion, a wriggly line a generation-changing gauge boson 
and a dashed line a generation-changing Higgs boson of the type $\phi^{(a)}$ 
detailed above.  Of these remaining diagrams, Figures \ref{1loopdiag}(a)
and (b) give rotations of order $m^2/\zeta_0^2$ (where $\zeta_0$ 
is the smallest Higgs vev) and are constrained by experiment to
be negligible for the following reason.  As noted before, in breaking 
the generation $SU(3)$ symmetry, the corresponding gauge bosons all
acquire masses of order or higher than $g \zeta_0$, $g$ being the 
gauge coupling.  The
exchange of these bosons will lead to flavour-changing neutral current
(FCNC) effects at low energies of the order $g^2/(g \zeta_0)^2 
\sim 1/\zeta_0^2$.  
Present experimental bounds on FCNC effects, such as an anomalous 
$K_L-K_S$ mass 
difference, will thus lead to very stringent lower bounds on the value of
$\zeta_0$, which is currently of the order 100 TeV \cite{ourFCNC}.  Hence
the rotation due to Figures (a) and (b), even for the top quark of mass
180 GeV, is only of order $10^{-6}$ and therefore entirely negligible.
There remains then only the Higgs loop diagram (c) to be considered.

The rotation from the diagram (c) has been evaluated \cite{ourCKM} and
gives:
\begin{equation}
\frac{d}{d(\ln \mu^2)} \left( \begin{array}{c} x' \\ y' \\ z' 
   \end{array} \right)
   =  \frac{5}{64 \pi^2} \rho^2 \left( \begin{array}{c}
         \tilde{x}'_1 \\ \tilde{y}'_1 \\ \tilde{z}'_1 \end{array} \right),
\label{runxyz}
\end{equation}
with
\begin{equation}
\tilde{x}'_1 = \frac{x'(x'^2-y'^2)}{x'^2+y'^2} + \frac{x'(x'^2-z'^2)}
   {x'^2+z'^2}, \ \ \ {\rm cyclic},
\label{x1tilde}
\end{equation}
and $\rho^2 = |a|^2 + |b|^2 + |c|^2$ being the Yukawa coupling strength.
By iterating this formula, one can compute the trajectory traced out by 
the vector $(x', y', z')$ given any initial value.

The choice of an initial value of the vector $(x', y', z')$, which fixes 
the trajectory it is on, depends in principle on the original vev's 
$x, y, z$ of the Higgs fields, the masses of the Higgs bosons, and also 
the Yukawa coupling strength $\rho$, the last of which depends in turn on the 
fermion-type.  One can thus attempt a global fit to the empirical CKM 
matrix and fermion mass spectrum with these quantities as parameters.
This was the approach adopted in \cite{ourCKM} and a good fit has 
been obtained.  In this paper, however, we shall consider only a particular
solution suggested by the fit in \cite{ourCKM} which we believe may have 
a deeper meaning than is as yet fully understood, namely when the Yukawa 
coupling strength $\rho$ is the same for all fermion-types.\footnote{Notice 
that the normalization of the mass matrix is not calculable perturbatively 
if the coupling is large as in the DSM scheme, and has thus to be regarded 
in general as a different parameter from the Yukawa coupling $\rho$ in the 
present framework.}  In this case, the vector $(x', y', z')$ runs on the
same trajectory with the same speed for all fermion-types which differ
thus only in the positions that their physical states occupy on the common
trajectory.  This simplifies the problem considerably and renders the 
mechanism very transparent since the whole set-up now depends on only 
3 (real) parameters, namely the common Yukawa coupling strength $\rho$ 
and a common (normalized) initial vector $(x_I, y_I, z_I)$ at some (high) 
arbitrary scale.  With these, as we shall see, one can already explain 
semi-quantitatively nearly all the features of quark and lepton mixing 
noted above, while making as well some rough estimates for the lower 
generation fermion masses given the masses of the heaviest generation.  

Before we proceed to a formal fit of the data with the 3 remaining 
parameters, let us first examine the problem qualitatively to try to
anticipate the form that such a fit will take.  From (\ref{runxyz}) and 
(\ref{x1tilde}), one sees that $(1,0,0)$ and $\frac{1}{\sqrt{3}}(1,1,1)$ 
are both fixed points on the trajectory, and that when going down in 
energy scale, the vector $(x', y', z')$ runs away from $(1,0,0)$ towards 
$\frac{1}{\sqrt{3}}(1,1,1)$.  It will run, of course, faster in the 
middle than near the fixed points, at a speed the actual value of which 
depends on the Yukawa coupling strength $\rho$.  

\begin{figure}[htb]
\vspace{-5cm}
\centerline{\psfig{figure=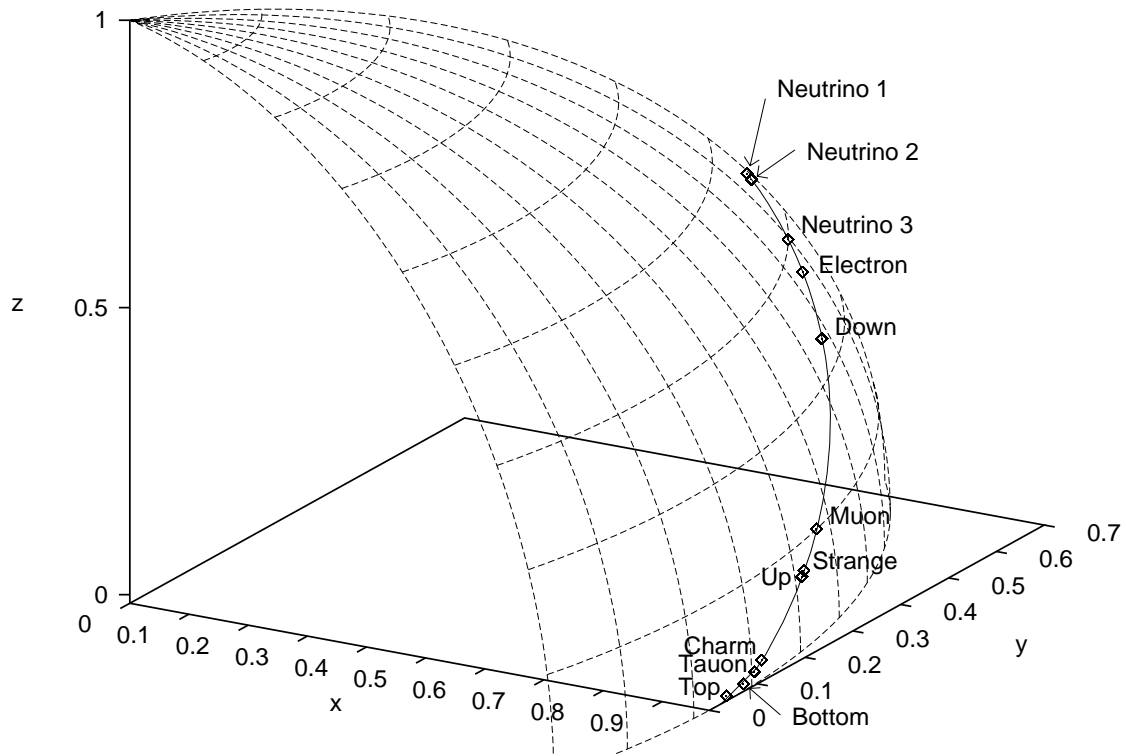,width=0.85\textwidth}}
\caption{Trajectory traced out by the vector $(x', y', z')$.}
\label{runtraj}
\end{figure}

Consider first the fermion masses of the two highest generations, where  
one recalls that in the present set-up masses of the second generation 
arise only by `leakage' from the highest generation.  It follows then 
from the observation in the above paragraph that those situated near 
the fixed points will acquire proportionately smaller masses from `leakage' 
since the running is there less efficient.  Given now the empirical pattern 
that $m_c/m_t < m_s/m_b < m_\mu/m_\tau$, while $m_t > m_b > m_\tau$, 
namely the heavier the mass the smaller the `leakage', it seems advisable 
in attempting a fit to place $m_t$ fairly close to the high energy fixed 
point $(1,0,0)$, so that $m_b$ and $m_\tau$ being lower in mass and hence
further away from the fixed point will `leak' more of their masses into 
their second generation states.  The resulting arrangement for the 2 highest 
generation states of the 3 fermion-types $U, D, L$ would then roughly be 
as shown in Figure \ref{runtraj}.  

 For neutrinos $N$, the consideration is a little more complicated.  
What enter in the `leakage' argument of Figure \ref{runtraj} are the 
Dirac masses $M_{\nu_i}$, but neutrinos can also have a Majorana mass 
$B$ \footnote{In order for the `leakage' mechanism to work for 
neutrinos as for the other fermion-types, they have also to be Dirac 
fermions with their left-handed components forming a triplet of the 
horizontal $SU(3)$ symmetry and their right-handed components $SU(3)$ 
singlets having a common Majorana mass.}.  The physical masses $m_{\nu_i}$
for the 3 generations of neutrinos are given by the see-saw mechanism 
as $M_{\nu_i}^2/B$.  Experimentally, if neutrino masses are assumed to be 
hierarchical, as they must be in the present set-up, the data on atmospheric 
neutrinos \cite{kamioka,superk,soudan} give a (physical) mass to the 
heaviest neutrino $\nu_3$ of order $m_{\nu_3}^2 \sim 10^{-3}-10^{-2} 
{\rm eV}^2$.  For the second generation neutrino $\nu_2$, solar neutrino 
data suggest a (physical) mass of either $m_{\nu_2}^2 \sim 10^{-5} {\rm eV}^2$ 
if one takes the MSW solution \cite{MSWfit}, or $m_{\nu_2}^2 \sim 10^{-10} 
{\rm eV}^2$ if one takes the LWO solution \cite{LWOfit}.  In the MSW 
case, one obtains then $M_{\nu_2}/M_{\nu_3} \sim 0.18-0.31$, while in
the LWO case $M_{\nu_2}/M_{\nu_3} \sim 0.010-0.018$.  This ratio for
the MSW case is much bigger than the corresponding figures for the other 
3 fermion-types $U, D, L$, which in the present set-up means also bigger 
`leakage efficiency'.  Indeed, the `leakage' required by the MSW solution 
is so big that one is easily convinced by a few trial calculations that it
cannot be accommodated here even if $\rho$ is allowed to take a very
different value from the other 3 fermion-types.  On the other hand, the 
`leakage efficiency' required by the LWO solution, which is only somewhat
bigger than that of the $U$-type quarks, can be readily accommodated.  Since 
the Dirac masses of neutrinos (dependent on $B$) are empirically unknown, 
the heaviest state $\nu_3$ can in principle be assigned any location on 
the trajectory so long as it gives a correct `leakage efficiency' to 
reproduce the mass ratio $M_{\nu_2}/M_{\nu_3}$.  One obvious possibility
is to locate $\nu_3$ close to $t$ but this will make the lepton CKM matrix
very similar to that of the quarks.  A much more interesting possibility 
is to place $\nu_3$ far down the same trajectory, as illustrated in Figure 
\ref{runtraj}, where since the vector $(x', y', z')$ is now pressing 
against the low energy fixed point $\frac{1}{\sqrt{3}}(1,1,1)$ the 
`leakage efficiency' is again reduced, say compared to $D$-type quarks 
and charged leptons, as required.  We choose to consider this second 
possibility. 

As explained above, giving the locations on the trajectory of the 2
highest generation states in the present set-up also fixes the triad
of state vectors of all 3 generations.  It is then a simple matter to 
evaluate the CKM matrix the elements of which are just the direction 
cosines between the triads of the $U$- and $D$-type quarks, or else for
leptons, between the triads of the charged leptons $L$ and the neutrinos
$N$.  Given that in Figure \ref{runtraj}, the quarks are much closer in
location than the leptons, so also will be their triads in orientation.
It follows then immediately (a) that the CKM matrix is much closer to the 
identity for quarks than for leptons, a qualitative fact clearly borne 
out by a comparison between the empirical CKM matrices (\ref{exckmq}) 
and (\ref{exckml}).  

To study further the details of the various elements, it is convenient 
to consider the limit when the locations on the trajectory of the 2 
highest generations are close together so as to make use of some
familiar formulae in elementary differential geometry.  This is seen in 
Figure \ref{runtraj} to be a reasonable approximation at least for the 3 
fermion-types $U, D, L$.  In this case, the triad of state vectors in 
Figure \ref{triad} becomes the so-called Darboux triad \cite{docarmo} 
with (i) ${\bf v_3}$ being the (radial) vector normal to the surface 
(sphere), (ii) ${\bf v_2}$ the tangent vector to the curve (trajectory), 
and (iii) ${\bf v_1}$ the vector orthogonal to both.  And the CKM 
matrix becomes just the rotation matrix of the Darboux triad on
transporting it along the trajectory from the $U$ to the $D$ location 
for quarks, or from the $L$ to the $N$ location for leptons.  To first
order in the displacment, this rotation matrix is given by a variant of
the well-known Serret-Frenet formula:
\begin{equation}
CKM \sim \left( \begin{array}{ccc} 1 & -\kappa_g \Delta s & -\tau_g \Delta s \\
                                   \kappa_g \Delta s & 1 & \kappa_n \Delta s \\
                                   \tau_g \Delta s & -\kappa_n \Delta s & 1
         \end{array} \right).
\label{darboux}
\end{equation}
 For our special case of a curve on a unit sphere the geodesic torsion
vanishes $\tau_g = 0$ and the normal curvature is constant $\kappa_n = 1$.  
As a result, one concludes immediately (b) that the corner elements of 
the CKM matrix, being of at least second order in the the displacement 
$\Delta s$, are much smaller than the others, and (c) that the 23 and 
32 elements, being proportional to the separation between $t$ and $b$ 
for quarks and between $\tau$ and $\nu_3$ for leptons, are much smaller 
for the quark than for the lepton CKM matrix.  Again, as already noted 
at the beginning, these predictions are strongly borne out by experiment.  
The other two off-diagonal elements depend on the geodesic curvature 
$\kappa_g$ which depends in turn on both the trajectory and the location 
on it, and will be harder for the present mechanism to predict.

One sees therefore that, even without performing any calculation, one is 
already able to explain qualitatively most of the outstanding features in 
the mixing pattern and the hierarchical mass spectrum of both quarks and 
leptons.  What remains now is to attempt an actual fit with our 3 parameters 
and see if one gets reasonable quantitative agreement.  We propose to proceed
as follows.  Of the quantities we can calculate, the most accurately 
measured experimentally are the 2 mass ratios $m_c/m_t, m_\mu/m_\tau$ and 
the Cabibbo angle $V_{cd} \sim V_{us}$.  We shall therefore determine
our 3 parameters by fitting the experimental values of these 3 quantities.
Having then decided on a trajectory for the (normalized) vector $(x',y',z')$
as encoded in some intitial value $(x_I,y_I,z_I)$, and on the value of the 
Yukawa coupling strength $\rho$ which governs the speed with which the
vector runs along the trajectory, we can then just follow the procedure
given above to calculate the other parameters.  We have to input the (Dirac)
masses of the heaviest generation.  For the $U$- and $D$-type quarks and
charged leptons, we take from \cite{databook}:
\begin{equation}
m_t = 173.8 \ {\rm GeV}, \ \  m_b = 4.247 \ {\rm GeV}, \ \ 
   m_\tau = 1.777 \ {\rm GeV},
\label{inputmassql}
\end{equation}
the chosen value for $m_b$ being the geometric mean of the given experimental 
limits.  With these inputs, we calculate the masses of $c$ and $\mu$ 
and the quark CKM matrix elements $V_{us}$ and $V_{cd}$, adjusting 
the values of the Yukawa coupling strength $\rho$ and the initial values 
of the vector $(x_I, y_I, z_I)$ until we obtain the experimental values 
given in \cite{databook}, namely:
\begin{equation}
m_c = 1.1 - 1.4 \ {\rm GeV},\ \ \  m_\mu = 105.6 \ {\rm MeV},\ \ \ 
   V_{us}, V_{cd} = 0.217 - 0.224.
\label{fitvalues}
\end{equation}
This requires running the vector $(x', y', z')$ numerically with the 
formula (\ref{x1tilde}) from the initial value $(x_I, y_I, z_I)$ down
to the second heaviest generation for each fermion-type.  We take typically
around 500 steps for each decade of energy to achieve about 1 percent 
accuracy, normalizing the vector $(x', y', z')$ again at every step.  
The quantities $\rho$ and $m_T$ which in principle also run 
are taken here, for lack of anything better, to be constants, any slow 
variations of which, we believe, would be masked in practice by adjustments 
of the free parameters to fit the values in (\ref{fitvalues}).  With the 
values of $\rho$ and $x_I, y_I, z_I$ so obtained, we can then make 
predictions for other quantities.

We distinguish two categories of such predictions.  The first requires
only the running between the heaviest and second heaviest generations which
category is expected to be more reliable given that our parameters have 
been determined from running in the same range.  These predictions include 
all the CKM matrix elements for both quarks and leptons, and the masses of 
the strange quark $m_s$ and the `right-handed neutrino' $B$.  A list of
such predictions on the CKM matrix elements is given in Table \ref{CKMtable}
where the `predicted central value' is obtained by putting $m_t = 173.8 
\ {\rm GeV}$, the experimental central value, $m_c = 1.241 \ {\rm GeV}$, the 
geometric mean of the experimental limits, and $\frac{1}{2}(V_{us}+V_{cd}) 
= 0.2205$, the (arithmetic) mean of the experimental limits, giving for the
`central values' of the fitted parameters:
\begin{equation}
\rho = 3.535, \ \ x_I = 0.9999984, y_I = 0.0017900, z_I = 0.0000179,
\label{fitparc}
\end{equation} 
where the initial value of the vector $(x_I, y_I, z_I)$ is taken arbitrarily 
at the scale of 20 TeV.  The `predicted range' is obtained by varying $m_t$ 
within the quoted experimental error of $\pm 5.2 \ {\rm GeV}$, and $m_c$ and 
$V_{us}, V_{cd}$ within their experimental limits quoted in (\ref{fitvalues}) 
above, and corresponds to the range of the fitted parameters:
\begin{equation}
\begin{array}{ll}
\rho  = 3.393 - 3.688, & x_I  = 0.9999959 - 0.9999994,   \\
   y_I = 0.0010800 - 0.0028500, & z_I  = 0.0000075 - 0.0000391.
\label{fitparr}
\end{array}
\end{equation}
The agreement between prediction and experiment for the quark CKM matrix
in Table \ref{CKMtable} is seen to be good for all entries.

 For neutrinos, as explained above, we need to input the physical masses 
of the two heaviest generations.  Taking these as:
\begin{equation}
m_{\nu_3}^2 = 3.5 \times 10^{-3} {\rm eV}^2, \ \ m_{\nu_2}^2 = 4.3 \times 
   10^{-10} {\rm eV}^2,
\label{inputmassnu}
\end{equation}
which are the best fit values to the latest SuperKamiokande data given in
\cite{newSK1,newSK2}, one obtains the entries for the lepton CKM matrix 
in Table \ref{CKMtable}.  On the other hand, if one varies these input
masses within the range permitted still either by \cite{kamioka,soudan} or
by \cite{superk,newSK1,newSK2}:
\begin{equation}
m_{\nu_3}^2 = (1.2 - 30) \times 10^{-3} {\rm eV}^2, \ \ 
   m_{\nu_2}^2 = (0.6 - 7.9) \times 10^{-10} {\rm eV}^2,
\label{inputnurange}
\end{equation}
while keeping the central values (\ref{fitparc}) of the fitted parameters,
one obtains:
\begin{equation}
U_{\mu3} = 0.6434 - 0.7108,\ \  U_{e3} = 0.0617 - 0.0814, \ \  
   U_{e2} = 0.2221 - 0.2352.
\label{Urange}
\end{equation}
The agreement with experiment is again seen consistently to be good, except 
for $U_{e2}$.  Notice in particular, by comparing with the quark matrix, 
the close agreement with the outstanding features (a)--(c) of the empirical 
mixing matrices noted at the beginning.  The element $U_{e3}$ is small as
required by \cite{chooz} while $U_{\mu3}$ responsible for the muon anomaly
in atmospheric neutrinos is near maximal corresponding to $\sin^2 2\theta > 
0.97$.  As for $U_{e2}$, the mixing angle involved in oscillations of solar 
neutrinos, we recall from (\ref{darboux}) above that, of the mixing elements 
in the CKM matrix, this element corresponding to the `geodesic curvature' 
$\kappa_g$ is the one most sensitive to details in the present scheme, being 
dependent both on the trajectory and on the location on it.  It is therefore 
not surprising that, though still of a reasonable order of magnitude, it 
does not come out as well as the others.  

In addition, one predicts:
\begin{equation}
m_s = 173 \pm 5 \ {\rm MeV}, \ \ \ B = 300\ (223 - 418) \ {\rm TeV}.  
\label{msandB}
\end{equation}
The value of $m_s$ given is the running mass taken at the scale equal to its 
value and cannot be directly compared with the values given in the data tables,
e.g. 100 - 300 MeV taken at 1 GeV \cite{databook96} or 70 - 170 MeV taken at
2 GeV \cite{databook}, but is seen to be reasonable.  The predicted value
for $B$, which is of course experimentally yet unknown, is interesting
in that it is much lower than usual GUT estimates and leads to much more
accessible rates for neutrinoless double beta decays, only 2--3 orders of
magnitude lower than the present limit.

\begin{table}
\begin{eqnarray*}
\begin{array}{||c||c||c|c||}  
\hline \hline
Quantity & Experimental Range & Predicted & Predicted Range \\
         &                    & Central Value &             \\
\hline \hline
|V_{ud}| & 0.9745 - 0.9760 & 0.9753 & 0.9745 - 0.9762 \\ \hline
|V_{us}| & 0.217 - 0.224 & (0.2207) &                 \\ \hline
|V_{ub}| & 0.0018 - 0.0045 & 0.0045 & 0.0043 - 0.0046 \\ \hline
|V_{cd}| & 0.217 - 0.224 & (0.2204) &                 \\ \hline
|V_{cs}| & 0.9737 - 0.9753 & 0.9745 & 0.9733 - 0.9756 \\ \hline
|V_{cb}| & 0.036 - 0.042 & 0.0426 & 0.0354 - 0.0508 \\ \hline
|V_{td}| & 0.004 - 0.013 & 0.0138 & 0.0120 - 0.0157 \\ \hline
|V_{ts}| & 0.035 - 0.042 & 0.0406 & 0.0336 - 0.0486 \\ \hline
|V_{tb}| & 0.9991 - 0.9994 & 0.9991 & 0.9988 - 0.9994 \\ \hline
|V_{ub}/V_{cb}| & 0.08 \pm 0.02 & 0.1049 & 0.0859 - 0.1266 \\ \hline
|V_{td}/V_{ts}| & < 0.27 & 0.3391 & 0.3149 - 0.3668 \\ \hline
|V_{tb}^{*}V_{td}| & 0.0084 \pm 0.0018 & 0.0138 & 0.0120 - 0.0156 \\ \hline
   \hline
|U_{\mu3}| & 0.56 - 0.83 & 0.6658 & 0.6528 - 0.6770 \\ \hline
|U_{e3}| & 0.00 - 0.15 & 0.0678 & 0.0632 - 0.0730 \\ \hline
|U_{e2}| & 0.4 - 0.7 & 0.2266 & 0.2042 - 0.2531 \\ \hline \hline 
\end{array}
\end{eqnarray*}
\caption{Predicted CKM matrix elements for both quarks and leptons}
\label{CKMtable}
\end{table} 

The other category of predictions requires running further down in energy
scale down to the lightest generation with parameters fixed by fitting the 
two heavier generations.  First, being extrapolations on a logarithmic 
scale, they are in any case not expected to be reliable except as rough 
order-of-magnitude estimates.  Secondly, for quarks, nonperturbative QCD
corrections are important below 1-2 GeV, which are hard to estimate.
Nevertheless if one persists, assuming still $\rho$ and $m_T$ to be 
constants, one obtains:
\begin{equation}
m_u = 200 \ {\rm MeV}, \ \ \ m_d = 15 \ {\rm MeV}, \ \ \ m_e = 6 \ {\rm MeV}, 
   \ \ \ m_{\nu_1} \sim 2 \times 10^{-15} \ {\rm eV}, 
\label{lightestp}
\end{equation}
to be compared with the experimental numbers:
\begin{equation}
m_u = 1.5 - 5 \ {\rm MeV}, \ \ \ m_d = 3 - 9 \ {\rm MeV}, \ \ \ 
   m_e = 0.51 \ {\rm MeV},  \ \ \ m_{\nu_1} < 10\ {\rm eV}.
\label{lighteste}
\end{equation}
While $m_d$ and $m_e$ may be considered reasonable given the expected
inaccuracy and $m_{\nu_1}$ has of course no difficulty in satisfying the
experimental bound, the predicted value for $m_u$ is some 2 orders out.  
It should be stressed, however, that the predicted value for $m_u$ is 
defined as the running mass taken at the scale equal to its value, and 
it is unclear whether it should be compared with the quoted experimental 
value defined at the scale of 2 GeV.  Indeed, if one simply calculates 
the expectation value in the $u$-state of the running mass matrix $m'$ 
at GeV scale, one obtains a value of order only 1~MeV, but it is also 
unclear whether this is the number to be compared to the quoted experimental 
value.  Barring this ambiguity, which applies also to $m_d$, the comparison
to experiment at an order-of-magnitude level is not unreasonable as the 
masses do at least follow the clear hierarchical pattern seen in experiment.  

One concludes therefore that simply by assuming that generations originate
in an $SU(3)$ gauge symmetry broken in the particular manner of (\ref{Vofphi}),
one can already explain the main empirical features in the mixing pattern 
together with the hierarchical mass spectrum of the Standard Model fermions.  
An important feature of the mechanism is that the mixing pattern and the
hierarchical mass spectrum are intimately related.  In particular, one 
recalls that for neutrinos, the mass ratio $m_{\nu_2}/m_{\nu_3}$ between 
the two heaviest generations cannot be as large as that required by the 
standard MSW solutions to the solar neutrino problem, or otherwise one finds 
no solution with the present mechanism, which admits only mass ratios of the
order of that required by the vacuum or long wave-length (LWO) solution.  
Hence, if the preference of the recent SuperKamiokande data for the (LWO) 
solution (iii) is maintained, it would lend support to this mechanism.

Further, one has recovered here the bulk of the phenomenological output of 
what we called the Dualized Standard Model (DSM) without having introduced 
at all the concept of nonabelian duality on which that scheme is based 
\cite{ymduality}.  The only phenomenological consequence of DSM so far 
studied which has been missed by the considerations here is the possible
explanation of cosmic ray air showers beyond the Greisen-Zatsepin-Kuz'min 
cut-off.  There seems thus a 
valid case to consider the present mechanism on its own independently of 
the original `dual' tenets of the DSM.  Indeed, one might attempt to go a
step further and bypass even the particular symmetry breaking scheme
embodied in the Higgs potential (\ref{Vofphi}), for the main effect
of that was really just to make the mass matrix factorize and rotate
with respect to the energy scale.  If one can devise some other scheme in 
which a similar situation attains, then an analogous conclusion is likely
to be achievable for explaining the empirically observed mixing pattern.

We ourselves, however, adhere to our preference for the original DSM scheme.  
The reason is that not only does the dynamical mechanism examined in this
paper arise naturally there as a consequence of the dual framework, but
even the very existence itself of a broken $SU(3)$ gauge symmetry and of
the Higgs fields required for its breaking emerges automatically from 
the concept of nonabelian duality.  Indeed, if one accepts this concept, 
then the niches for `generations' and `Higgs fields' would in any case 
already exist in the Standard Model, and if they are not assign these 
their seemingly natural physical roles, they would still have to be 
accounted for in some other manner, which may not be easy to come by.

Lastly, it should be stressed that although the main features of fermion
mass and mixing patterns are shown to follow from the dynamical mechanism
described in this paper, no consideration has been given here for possibly 
other predictions of the same mechanism violating experiment.  For the
DSM scheme, some considerations have been given to these questions, but if
this mechanism is grafted on to some other specific scheme, such questions 
will of course have to be readdressed.

\vspace{.5cm}

\noindent {\large {\bf Acknowledgement}}

\vspace{.2cm}

One of us (JB) is supported in part by grants CYCIT96-1718, PB97-1261 
and GV98-1-80.  He would also like to thank the Rutherford Appleton
Laboratory for hospitality.

\end{document}